# COMBINING PIN AND BIOMETRIC IDENTIFICATIONS AS ENHANCEMENT TO USER AUTHENTICATION IN INTERNET BANKING

*Cherinor Umaru Bah[1,*], Afzaal Hussain Seyal[1], Umar Yahya[2]*
{M20171042@student.utb.edu.bn, afzaal.seyal@utb.edu.bn, umar.yahya@ieee.org}

[1]*School of Computing and Informatics, Universiti Teknologi Brunei, Brunei*
[2]*Motion Analysis Lab, Faculty of Science, Universiti Brunei Darussalam, Brunei*
*Corresponding author

## Abstract

Internet banking (IB) continues to face security concerns arising from illegal access to users' accounts. Use of personal identification numbers (PIN) as a single authentication method for IB users is prone to insecurities such as phishing, hacking and shoulder surfing. Fingerprint matching (FPM) as an alternative to PIN equally has a downside as fingerprints reside on individual mobile devices. A survey we conducted from 170 IB respondents of 5 different banks in Brunei established that majority (65%) of them preferred use of biometric authentication methods. In this work, we propose a two-level integrated authentication mechanism (2L-IAM). At the first level, the user logs in to their IB portal using either PIN or FPM. At the second level, user is authenticated by means of face recognition (FR) should they initiate a transaction classified as sensitive. The merits of the introduced 2L-IAM are 3-fold: - (1) FR guarantees the identity of the rightful user irrespective of the login device; (2) By classifying banking products' sensitivity, the sensitive transactions are more effectively secured; (3) It is accommodative of different users' authentication preferences. Adoption of this framework could thus improve both users' and banks' experiences in terms of enhanced security and service delivery respectively.

**Keywords:** Internet Banking; User Authentication; PIN; fingerprint; face recognition

## 1 Introduction

Internet Banking (IB) has grown tremendously over the years as banking institutions strive to compete in a growing technological age in which customer convenience, security and cost minimization are among the crucial success factors[1-2]. In a typical IB portal, authenticated users can view their account balances, retrieve a list of recent transactions, transfer funds between accounts, pay utility bills, among several other online transactions[3]. It is, however, a growing challenge that IB continues to face evolving security challenges characterised by data compromise, phishing, hacking, and several more [4]. These prevalent challenges require banking institutions to continuously work towards the enhancement of user authentication mechanisms in IB environments[5]. The decades' long means of authentication such as passwords, personal identification numbers (PINs), and tokens no longer are able to keep up with the advanced tools and techniques used by hackers and fraudsters to gain illegal access to customers IB portals[6-8].

Biometric authentications which verify users based on their physiological attributes such as finger-prints, face and voice recognition, and behavioural characteristics such as, gesture, keystrokes, and gait are increasingly becoming the preferred authentication mechanism today [9-11]. IB and financial technology experts have even predicted that in the future, biometrics will replace the traditional electronic payment authentication method [12-13].

Basing on both the strength and weaknesses of both biometric and non-biometric authentications [6-14], we hypothesize that using a biometric fingerprint as well as facial recognition in combination with the other traditional methods such as passwords, PINs, and tokens would outperform the existing user authentication methods. Our hypothesis is on the basis that biometrics too possess some limitations which sometimes make the performance of one single modality insufficient in terms of universality, accuracy and distinctiveness [14-16]. In order to overcome these limitations, multi-biometric is increasingly attracting the interest of researchers with the main purpose of overcoming mono-modal biometric systems [17].

This work proposes a two-level integrated authentication mechanism (2L-IAM) that will use Personal Identification (PIN) or Fingerprint (FP) at the first level and at the second level the user is authenticated by means of Face Recognition (FR) should they attempt to perform any actions or transactions classified as sensitive by their bank. The rationale of the proposed 2L-IAM is that a second level authentication by means of FR guarantees the true identity of the rightful IB user. Additionally, banks would be able to decide and effectively categorise online transactions requiring additional authentication before being executed by their IB users. Therefore it was anticipated that the proposed 2L-IAM would offer enhanced user experience for both IB users and service providers (i.e banks) in the form of improved security, flexibility, recognising user authentication preferences and convenience as highlighted in [2-5].

# 2 Methods

## 2.1 Survey conducted

A survey of 170 IB users (aged 25-50 years) drawn from 5 major banks in Brunei Darussalam was conducted to establish their preference between PIN and biometric authentication methods. The survey aimed to reconfirm from a primary source three main aspects with regards to IB experience. The aspects included; (i) Establishing user preference with respect to IB and the traditional physical branch-based banking, (ii) To establish the extent of use and rating of user experience in using PIN, tokens, and biometric authentication methods (BAM), and (iii) To establish user preference between BAM and other authentication mechanisms. A combination of findings from this survey and the evidence on users' IB experience established from literature [3-5, 18-19] formed the basis for the 2L-IAM proposed in this study.

## 2.2 Proposed User Authentication Framework

The components (layers) of proposed 2L-IAM are illustrated in Fig.1. Layer 1 comprises of the end-user application interface platforms. Layer 2 and layer 4 comprise of the first-level and second-level authentication modules, respectively. Layer 3 represents the IB knowledge base module which comprises of the banking database, internet infrastructure, and the local and cloud storage of FP and FR files respectively.

The prototype was implemented using a combination of different scripting languages (i.e PHP, JSP, HTML, and SQL) interfaced with MySQL database server. Dreamweaver CS5 and Apache II web-server software were also used.

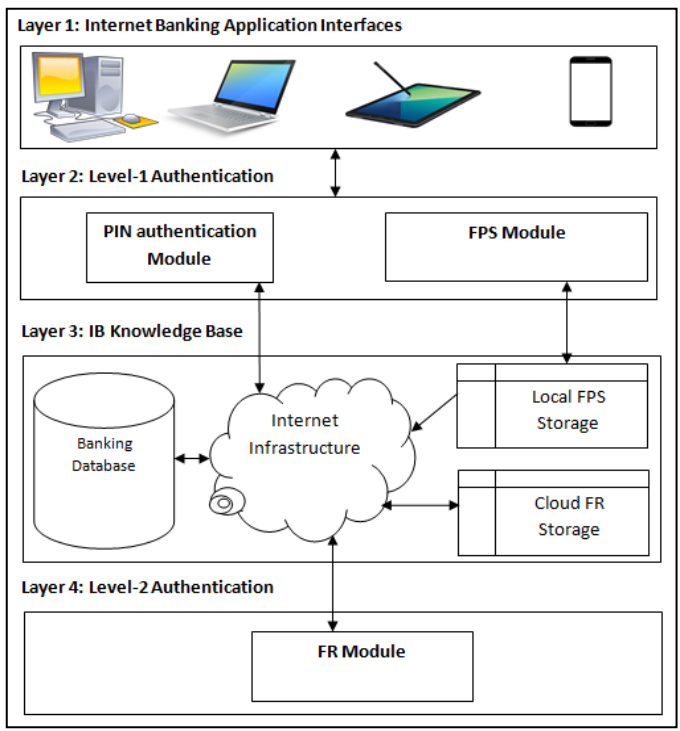

Figure 1: Components of the proposed 2L-IAM.

### 2.2.1 Application Interfaces

The prototype of 2L-IAM developed is accessible through any web browser and user's login options extend to any internet-enabled device (i.e laptop, mobile phone, tablet, PC, etc). The authentication module is also able to detect the type of device used and stored into the banking database.

### 2.2.2 Authentication Modules

*Level-1 Authentication (A1):* At this stage the IB user is authenticated by means of either a PIN (password) or FP scan (FPS) depending on the type of login device used. Users of smart mobile phones and tablets can use either of the two options, while desktop and laptop computer users are authenticated by PIN at this stage. FPS verification is achieved by means of matching the scanned FP against the stored one in the local FPS storage for a given device. PIN authentication is achieved by comparing the provided one against the existing one in the banking database via secure hypertext transfer protocol (Secure HTTP). In either case, the respective field in the user log of the banking database which identifies the type of authentication used is updated accordingly, once A1-access in granted.

*Level-2 Authentication (A2):* If the user with A1-access granted initiates a sensitive transaction (ST) as defined by the bank or as set by the user himself, the user is prompted to verify their true identity by means of FR. We defined ST in this work as those transactions deemed very crucial by the bank to a level requiring second-level verification before being executed. The prototype also proposes an option for a user to upgrade the status of any of the services to require A2 verification before being executed. However, users are not able to downgrade any of the services to A1 which are defined as A2 services by the bank. FR is achieved by means of matching the device-camera captured face of the user with that of the same user stored in the banking database in the cloud (i.e Cloud FR storage). The respective fields in the user log table of the banking database are updated successfully upon A2 being granted. Table1 shows the three possible statuses (S-1, S-2, and S-3) of a user at any given time as reflected in the banking database user log table.

Table 1: The three possible statuses of a user at any time

| Session Status | A1 | A2 | Description |
|---|---|---|---|
| S-1 | 0 | 0 | User is Offline |
| S-2 | 1 | 0 | User is Online |
| S-3 | 1 | 1 | User is Online and in Sensitive Mode |

### 2.2.3 IB Knowledge Base (IB-KB)

IB-KB is comprised of the banking database, local FPS storage, and Cloud FR storage. The banking database contains four main tables:- (i) the *user-details* table which stores IB user details including fields indexed to FPS and FR file systems, (ii) the *bank-services* table which among other attributes identifies services as either being A1 or A2 services, (iii) the *transactions* table which stores all previous effected transaction for respective services, and (iv) the *user-logs* table which stores all details pertaining to a specific IB user logon session including A1 and A2 time-stamps, type of device used, the geographical location of the user at the time of logging into the IB banking portal. Interfacing of the banking database with FP, PIN, and FR during user authentication and IB transactions occurs over Internet infrastructure through Secure HTTP as illustrated in Fig.1.

## 3 Results and Discussion

### 3.1 Testing of the developed 2L-IAM Prototype

Testing of the proposed 2L-IAM was performed for combinations of PIN and FR (Pair-1) as well as FPS and FR (Pair-2). Fig.2 illustrates access levels granted by A1 and A2 using Pair-1 authentication option.

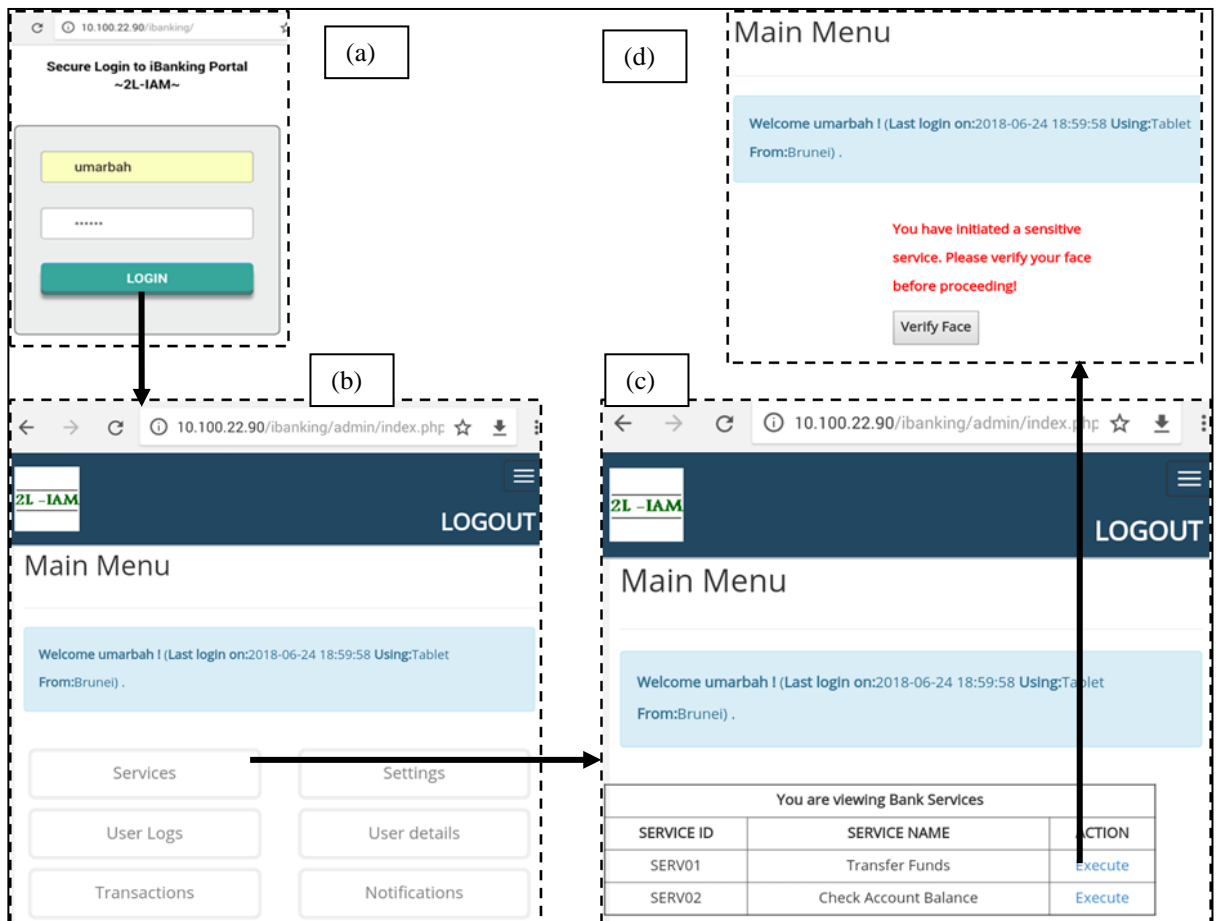

Figure 2: 2L-IAM Prototype Testing.

In (Fig.2a), a user is authenticated by means of PIN and once successful, access level A1 is granted where all the portal's functionalities are displayed (Fig.2b). When the user initiates a sensitive transaction from the list of bank's offered services (Fig.2c), the portal notifies the user of the second-level authentication required before proceeding (Fig.2d). Once the user's identity is re-verified by the FR authentication module, access level A2 is granted to allow execution of the transaction. The status of the IB user is updated in the banking database accordingly as described in Table 1.

### 3.2 Results of the Survey Conducted

The survey conducted in this work revealed four main findings: - (i) 59% of total respondents preferred IB while the rest still prefer traditional branch-based banking, (ii) 54% of total respondents reported that they find use of PIN, tokens, passwords to be a tasking option, (iii) 56% of total respondents reported fear for fraud in all their IB transactions for which the mode of authentication was non-BAM, and (iv) 65% of total respondents preferred use of BAM in IB while the rest still held preference for non-BAM.

### 3.3 Discussion

The survey findings of this study are in agreement with previous studies which report IB users' preference for IB to traditional branch-based banking, as well as preference for BAM to non-BAM [3, 6-13]. However, preference for BAM does not negate the reported fears of IB users with regards to storage of user biometrics by banks [14-20]. The proposed multi-factor authentication approach is a fair trade-off between security, convenience and user preferences. This is achieved by ensuring FP are only locally stored on users' mobile devices while allowing only PIN and FR storage onto the banking database in the cloud.

FR-based level-2 authentication in 2L-IAM is a more effective alternative to the currently used verification mechanisms by banks such as texting one-time use codes to user's registered phone number [18]. It is a better time-security trade-off than the text message option considering that the registered mobile phone might be switched off or in possession by an illegal user. Moreover the hardware requirement (i.e web camera / front-facing camera) for FR-based authentication is no major obstacle [20] considering that majority IB users currently use camera-enabled devices for IB. Adoption of this new proven concept would ideally require prior large-scale evaluation over a variation of factors such as web-camera quality, speed of internet connection, FPS technology used in mobile and tablet devices among several others.

## 4  Conclusions and Future Work

Basing on the findings of a primary survey conducted from 170 IB users of five major banks in Brunei Darussalam, and from existing literature on IB, a multi-factor 2L-IAM has been introduced in this work. The proposed authentication framework offers improved IB experience to both IB users and to banks offering IB services in the form of enhanced security, and better trade-off between security, convenience, and user authentication preferences. Future extensions of the proposed framework could explore use of various data mining techniques with the banking database in an attempt to automate and dynamically determine the type of second-level authentication to be enforced by the IB portal.

## Acknowledgements

Authors would like to thank all respondents from Brunei Darussalam who participated in the primary survey conducted in this work.